%% file: sample-authordraft.tex
\documentclass[10pt]{article}

\usepackage[letterpaper,top=0.8in,bottom=0.9in,left=0.9in,right=0.9in]{geometry}
\usepackage{times}       % main text font
\usepackage{mathptmx}    % math font matching Times
\usepackage{tabularx}
\usepackage{multirow}
\usepackage{booktabs}
\usepackage{graphicx}
\usepackage{amsmath}
\usepackage{amssymb}
\usepackage{url}
\usepackage{hyperref}
\usepackage{caption}
\usepackage[numbers]{natbib}
\AtBeginDocument{%
  \providecommand\BibTeX{{%
    \normalfont B\kern-0.5em{\scshape i\kern-0.25em b}\kern-0.8em\TeX}}}

\begin{document}

\title{Is Crowdsourcing a Puppet Show? Detecting a New Type of Fraud in Online Platforms}

\author{
  Shengqian Wang\\
  Ontario Tech University\\
  Oshawa, Ontario, Canada\\
  \texttt{shengqian.wang@ontariotechu.net}\\
  \texttt{ORCID: 0009-0009-2423-7305}
  \and
  Israt Jahan Jui\\
  Ontario Tech University\\
  Oshawa, Ontario, Canada\\
  \texttt{israt.jui@ontariotechu.net}\\
  \texttt{ORCID: 0000-0002-3644-3837}
  \and
  Julie Thorpe\\
  Ontario Tech University\\
  Oshawa, Ontario, Canada\\
  \texttt{Julie.Thorpe@ontariotechu.ca}\\
  \texttt{ORCID: 0000-0002-6629-158X}
}

\date{}
\maketitle

\begin{abstract}
Crowdsourcing platforms such as Amazon Mechanical Turk (MTurk) are important tools for researchers seeking to conduct studies with a broad, global participant base. Despite their popularity and demonstrated utility, we present evidence that suggests the integrity of data collected through Amazon MTurk is being threatened by the presence of \emph{puppeteers}, apparently human workers controlling multiple \emph{puppet} accounts that are capable of bypassing standard attention checks. If left undetected, puppeteers and their puppets can undermine the integrity of data collected on these platforms. This paper investigates data from two Amazon MTurk studies, finding that a substantial proportion of accounts (33\% to 56.4\%) are likely puppets. Our findings highlight the importance of adopting multifaceted strategies to ensure data integrity on crowdsourcing platforms. With the goal of detecting this type of fraud, we discuss a set of potential countermeasures for both puppets and bots with varying degrees of sophistication (e.g., employing AI). The problem of single entities (or puppeteers) manually controlling multiple accounts could exist on other crowdsourcing platforms; as such, their detection may be of broader application. 

While our findings  suggest the need to re-evaluate the quality of crowdsourced data, many previous studies likely remain valid, particularly those with robust experimental designs. However, the presence of puppets may have contributed to false null results in some studies, suggesting that unpublished work may be worth revisiting with effective puppet detection strategies.
\end{abstract}

\section{Introduction}
\begin{figure}
    \centering
    \includegraphics[width=0.7\linewidth]{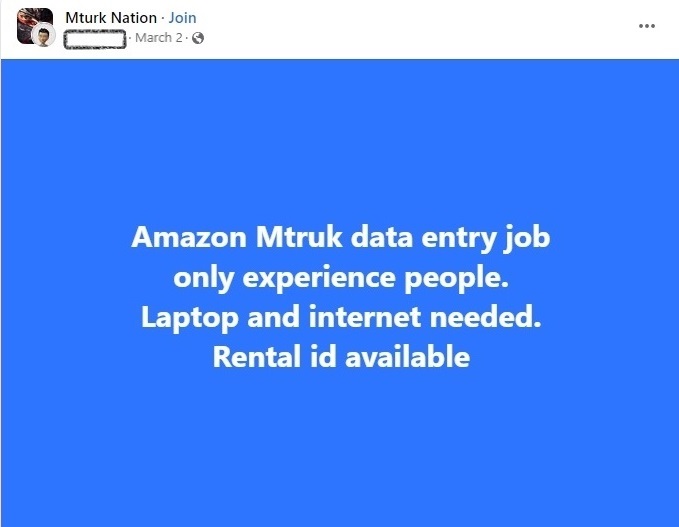}
    \caption{A screenshot for an advertisement related to rental Amazon MTurk accounts on Facebook, (captured on April 26, 2024.)}
    \label{fig:rental-MTurk}
\end{figure}

\begin{figure*}[tb]
\centering
  \includegraphics[width=0.8\textwidth]{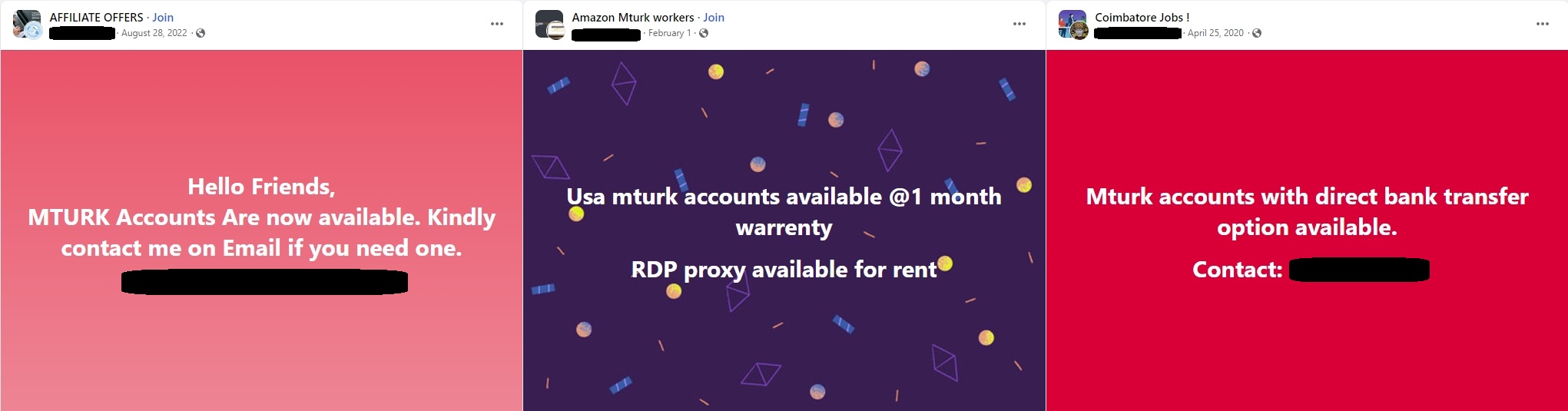}
  \captionsetup{width=.9\textwidth}
  \caption{Screenshots for Facebook posts related to Amazon MTurk trading in public groups, (captured on April 25, 2024).}
  \label{img:MTurk}
\end{figure*}

Crowdsourcing platforms such as Amazon Mechanical Turk (MTurk) have been popular among researchers as a versatile method of conducting user studies \cite{Chandler2016MTurkUsage}. However, Amazon MTurk has raised significant concerns in recent years, particularly regarding the reliability of data collected, as the growing amount of low-quality data has become a critical challenge \cite{hauser2019common,ternovski2022note}. Many researchers have investigated this issue, and have concluded that low-quality data comes from fraudulent workers \cite{Hauser2022CloudResearch}. Fraudulent workers frequently employ virtual private servers (VPS) or Virtual Private Network (VPN) to bypass geographical/IP address checks and enter random or illogical responses \cite{DennisMturk}. Some even buy established Amazon MTurk IDs with outstanding eligibility criteria (e.g., 95\% approval rates, and 5000 Human Intelligence Task (HITs)) from social media platforms like Facebook groups. In those groups, Amazon MTurk accounts are available in a variety of locations, such as the United States or India (see example posts in Figures \ref{fig:rental-MTurk} and \ref{img:MTurk}). Amazon MTurk workers who hold multiple accounts, or share their accounts with others for economic gain, violate Amazon MTurk Participation Agreement \cite{ParticipationAgreement}. Using the United States and Canada ``Identity Theft ByLaws'' as references \cite{USIdentityTheft,CanadaIdentityTheft}, this kind of activity can be classified as identity fraud. The continued activity of these Facebook groups, serves as a strong sign that Amazon MTurk's qualification filters are no longer effective, significantly impacting the Amazon MTurk's reliability. Consequently, researchers have put forward various suggestions to mitigate the issue of low-quality data. These suggestions include adding attention questions, cultural checks, using fingerprint techniques or employing combined statistical methods \cite{hauser2019common, Hauser2022CloudResearch}. These recommendations primarily operate under the assumption that each fraudulent worker owns his accounts, and works independently. Although Amazon can permanently ban their accounts, and researchers and other systems can blacklist known fraudulent accounts, fraudsters can still rent other qualified Amazon MTurk accounts from social media platforms, shown in Figure \ref{fig:rental-MTurk}.

\noindent\textbf{Contributions.} We shed light on a new type of fraud we observed in two user studies (N=558, N=689) using Amazon MTurk. Our contributions and findings can be summarized as follows: (1) We find that a significant source of noise on Amazon MTurk is a set of fraudulent workers (\emph{puppeteers}) who control multiple accounts (\emph{puppets}). In one study (N=558), 33\% of worker accounts exhibited this behaviour, while in a second study (N=689), the percentage could be as high as 56.4\%. We analyzed the UI interactions of these accounts for signals they may leverage automated programs (bots). Interestingly, our analysis does not support that these accounts are employing bots, but supports that these accounts have a human interacting with the system. 

(2) It is evident that employing a single approach is inadequate to address the challenge of fraud in crowdsourcing platforms. For example, attention checks can be easily evaded by these puppets if a human is behind them. Therefore, we discuss several countermeasures for researchers to establish their own effective checks, in order to minimize financial and labor costs.

\noindent\textbf{Impact.} This work highlights a new paradigm for understanding and detecting fraud in crowdsourcing platforms. Such fraudulent accounts compromise the integrity of data collected by them, and as such, is a security problem. This security problem has broad, and possibly unseen, impacts on research (and other) data across many disciplines. As such, detecting this type of fraud should serve a widespread benefit. Some subset(s) of detection methods may be applicable to other online platforms that can be negatively impacted by fraudulent accounts.

\section{Related Work}

Amazon MTurk is one of the most popular crowdsourcing platforms for researchers. It has been used for data collection services in a variety of domains \cite{Chandler2016MTurkUsage, bohannon2016psychologists}. Using such crowdsourcing platforms, researchers can easily register as task requesters, publish batches of questionnaires, and draw from a diverse global participant (``workers'') pool to gather valuable results \cite{Paolacci2014, mullinix2015generalizability}. This shift reflects a broader historical trend in many research fields toward leveraging technology to enhance research methodologies \cite{bhattacherjee2012social, chandler2019online, hunt2019using}. However, as with any methodological innovation, it brings its own set of challenges and considerations. Unfortunately, the quality from such crowdsourcing platforms has been dramatically reduced in recent years \cite{douglas2023data,schild2021behavior,Hauser2022CloudResearch,bohannon2016psychologists}, as a number of random, irrelevant, or identical responses (also defined as low quality data) have been observed ~\cite{buchholz2011crowdsourcing,gadiraju2015understanding,Florian2020Crowdsourced}. Participants who generate such responses are also referred to as ``fraudulent'' workers \cite{difallah2012mechanical,difallah2018demographics,eickhoff2013increasing,gadiraju2015understanding,schild2021behavior,kennedy_clifford_burleigh_waggoner_jewell_winter_2020}, or ``bots'' when they are automated programs rather than humans interacting with the system. 

This erosion of data quality raises concerns about the reliability of empirical research conducted through these platforms. The integrity of data is foundational to scientific knowledge, and compromised data can lead to invalid conclusions, echoing historical instances where data falsification undermined scientific progress ~\cite{fanelli2009many}. 

One effective method for obtaining high-quality data is to implement stringent eligibility filters for participants, such as a high number of prior completed HITs and approval rates. Dupuis et al.~ \cite{dupuis2022crowdsourcing} explored data quality issues across two studies on Amazon MTurk, each with different participant eligibility requirements. The first study, which had lower criteria (participants required at least 50 prior HITs with a 95\% approval rate), yielded a lower rate of usable responses at 12.4\% (N=429). In contrast, the second study, with more stringent criteria (participants needed a minimum of 1000 prior HITs with a 98\% approval rate), demonstrated a higher usable response rate of 31.3\% (N=342). Despite the improved response rate in the second study, the findings highlighted that a significant number of participants with higher eligibility still submitted low-quality responses or utilized tools to automate their answers.

This phenomenon can be examined through social science theories on human behavior and incentives. According to rational choice theory \cite{hechter1997sociological}, individuals are motivated to maximize their gains while minimizing effort. The economic incentives offered by researchers will encourage fraudsters to develop an automated system or work as groups to maximize their earnings with minimal effort. Furthermore, regulatory guidelines always require researchers to compensate participants regardless of the quality of the data they provide, which make fraudsters have no concerns when they submit irrelevant responses.

Another frequently recommended approach to addressing inattentive workers or bots issues involves the incorporation of attention questions, designed to assess a worker's engagement within the study \cite{hauser2016attentive, aguinis2021mturk}. These questions often involve simple mathematical calculations, recognition tasks. Workers who fail to select the correct option or challenges are flagged as inattentive workers, and their data is subsequently excluded from the research pool. While attention questions may help in filtering out certain ineligible workers and bots, they can be answered correctly by puppet accounts that are controlled by a puppeteer. 

In 2018, an Amazon MTurk ``bots'' crisis raised concerns among researchers who discovered a significant number of low-quality responses originating from the same geolocation \cite{Bai2018MTurkBots}, undermining the integrity of their databases \cite{Chandler2020MTurkBots, Dreyfuss2018MTurkBots, Stokel-Walker2018MTurkBots, Dennis2019OnlineWF, MargaretMturkBots}. CloudResearch, a platform that provides tools and services for conducting online research \cite{cloudresearch}, conducted two user studies to investigate the factors contributing to low-quality data production on Amazon MTurk in the United States, whether they were human or bots. They integrated attention questions and reCAPTCHA to their studies. The findings indicated that all of them were indeed humans when they passed those attention questions. Furthermore, their responses often displayed a lack of relevance to US culture, suggesting that these workers might be using VPS to bypass geolocation/IP address restrictions and doing the tasks manually.

Methods to collect participant’s unique information to identify if participants are bots, such as browser information, IP address, mouse and keyboard activities have been widely adopted by researchers \cite{marshall2023broke, buchholz2011crowdsourcing,chmielewski2020mturk,wood2017response}. However, those methods do not work efficiently, and can be evaded when participants disabled JavaScript, use VPN or VPS, or create false positives when the target participants are legitimately on the same network (e.g., students on campus or employees in the same building). Privacy concerns also arise when sensitive data such as IP addresses are collected, as improper storage and handling data procedures from researchers or organizations could accidentally leak this data.

Douglas et al. \cite{douglas2023data} examined the integrity of data collected from five online platforms: Amazon Mechanical Turk, Prolific, CloudResearch, Qualtrics, and SONA, an online global participant pool for undergraduate students \cite{SONA}. The assessment of data quality hinged on the participants' attentiveness and genuineness in responding, gauged through a series of attention checks and engagement metrics, including meaningful answers, memory of previously presented information, unique IP addresses and geolocations, and working at a pace that suggests they read all items. Remarkably, the study found that Prolific and CloudResearch recorded a commendable high-quality respondent rate of 67.94\% and 61.98\%. In contrast, Amazon MTurk and Qualtrics presented lower high-quality respondent rates of 26.40\% and 53.22\%. The significant number of low-quality responses are a strong indicator that some amount of fraudulent workers may exist across different platforms, not only on Amazon MTurk. Gadiraju et al. \cite{gadiraju2015understanding} conducted a user study of 1000 participants from Crowdflower, and examined the issue of fraudulent activities in a survey. They analyzed the behavior of untrustworthy workers, and classified them into 5 categories based on their behaviors: (1) Ineligible workers, who do not meet the prerequisite conditions set for the task but still attempt to participate. (2) Fast deceivers, who aim to complete tasks as quickly as possible, often resorting to providing irrelevant or copied responses, to earn rewards with minimal effort. (3) Rule breakers, who ignore specific task instructions or requirements, leading to responses that may not fully comply with the task's demands. (4) Smart deceivers, who are more calculated in their approach to deception, carefully crafting their responses to bypass checks and validations without triggering alarms, despite their responses being of poor quality or irrelevant. (5) Gold standard preys, who tried to provide useful responses but failed simple attention-check questions or certain requirements. The authors suggested several guidelines for survey design, such as employ pre-screening mechanisms to ensure that only eligible workers can participate in tasks, designing tasks to determine specific types of malicious behavior, and utilizing post-processing steps to identify responses from participants who may have been unfairly labeled as untrustworthy due to attention checks. 

Marshall et al. \cite{marshall2023broke} presented a detailed analysis of the declining reliability of data from (Amazon MTurk) from 2013 to 2022. They found unusable data surged from 2.4\% to 88.8\% over the years, and the effectiveness of traditional reliability checks like attention checks has deteriorated when participants became ``smart'' to pass those simple checks. They are the only other work we have seen that suggested that multiple account issues may exist, when a participant may collaborate or control several accounts when they have similar responses; however, they only raised the question of this possibility regarding two small ``clusters''. 

Our work builds upon these findings by providing compelling evidence of the puppeteer issue, quantifying its extent, and discussing detection techniques. This aligns with the historical progression of scientific inquiry, where new challenges prompt methodological advancements. Addressing the puppeteer issue is essential not only for individual studies but also for the broader credibility of research relying on crowdsourced data.

\section{Evidence of the Puppeteer Threat}

After analyzing the data from two separate Amazon MTurk studies, we observed some unusual patterns that suggested many Amazon MTurk users who participated were actually puppets. In this section, we briefly describe the studies, and present an analysis of the relevant data that suggests that the accounts were controlled by the same entity, and present an analysis of a number of indicators that there are likely humans interacting with the system (rather than bots). 

\subsection{Data Collection Methodology}

We conducted two user studies on Amazon MTurk in 2022, each intending to study a new authentication system. Neither was anticipating the issue of puppets, but they both had sets of suspicious results that led us to suspect their presence. Our University's Research Ethics Board approved both studies initially, and also for the secondary use of data for the analysis we present herein.

\subsubsection{User Study 1}

The objective of our initial user study, comprising 558 participants ($ N = 558 $) with (HIT $ \geq 95\% $, minimum completed 500 tasks, United States residents only), was to evaluate the performance of our password creation system. Participants were instructed to register an account, login, and answer questionnaires. There were three groups in total: one control, and two experimental. For participants in the experimental groups, they were shown and asked to interact with randomly generated content prior to password selection. Participants in the control group were only asked to create a password as in a typical password creation scenario.
 We carefully collected data on each participant's password selections, their responses to various questions, and their digital mouse and keyboard interactions, including timestamps of button clicks and scrolling activities, all of which were recorded within our system. 

The study involved two sessions using a system designed to inspire the creation of unique passwords:
\begin{itemize}
    \item Session 1 (5--10 mins, \$0.85 compensation). Each participant was asked to create a unique password and complete a questionnaire. Those who successfully completed the first session were invited back to log in to their accounts and fill out a second questionnaire.
    \item Session 2 (1--2 mins, \$0.35 compensation). The participant was asked to login with their password created in Session 1, then complete a questionnaire.

\end{itemize}

\subsubsection{User Study 2}

In a second independent study, comprising 698 participants ($ N = 698 $) with (HIT $ \geq 95\% $, minimum completed 500 tasks, United States residents only), we investigated if different methods can help and train users to memorize system-assigned 4-digit PINs on mobile devices. Participants were randomly allocated to one of four distinct conditions (one control, three involving different types of training), with each participant being assigned a unique PIN. The system worked by randomly generating a PIN on the server side, and storing it in the browser's local storage so that if the browser session exited by accident, a new PIN would not be assigned when the user returns. An unintended, but interesting consequence of this design choice was that if the user returned to the system using the same browser but a different Amazon MTurk ID, and was assigned by chance to the same group as the previous Amazon MTurk ID, their PIN would be identical. 

The study involved two sessions using involving a 4-digit PIN system:
\begin{itemize}
    \item Session 1 (5--8 mins, \$0.60 compensation). If the participant was in an experimental condition, they were asked to complete a short ($<1$ minute) training session. The control group was simply shown their assigned PIN. Then they were asked to login with their assigned PIN. After login, participants were asked to complete a short questionnaire. Those who successfully completed the first session were invited to Session 2.  
    \item Session 2 (1--2 mins, \$0.60 compensation). The participant was asked to login with their PIN assigned in Session 1, then complete a questionnaire.

\end{itemize}

\subsection{How To Determine An Account Could be a Puppet?} 

Study 1 and Study 2 each have unique data of interest to analyze for this purpose.  We take different approaches to determining an account is a puppet in each in their corresponding subsections.

\input{Study1}

\subsubsection{User Study 1}

While preparing Study 1's data for analysis, we observed numerous participants using identical passwords, some of which were notably unique. For each password observed more than once, we compute the probability ($P$) that the password appeared that many times (at least $k$ times) by chance. We utilized the binomial distribution for this purpose as follows. The general formula for calculating the probability that at least \(k\) participants out of \(n\) choose the exact same password is given by:

\begin{equation} \label{eq:1}
P(X \geq k) = 1- P(X < k) = 1 - \sum_{x=0}^{k-1} \binom{n}{x} \,p^x (1-p)^{n-x}
\end{equation}

\vskip 2mm

    To estimate \(p\), we retrieve the total prevalence counts of leaked passwords, denoted as $t = 5,579,399,834$, from the Pwned Passwords dataset \cite{PwnedPasswords}. Then we searched for the frequency of each password's occurrence ($o$) in this dataset. Then \(p = \frac{o}{t}\).

As shown in Table \ref{tab:probTable}, we found 31 distinct puppeteers, each having between 2--57 accounts.

\input{Study2}

\subsubsection{User Study 2}

Study 2 had a very straightforward way of detecting a potential puppet: whether the PIN number was the same for each account. When a user who has participated in the study returns, the system retrieves their previously-set random PIN from local storage. If in Session 1, there is no PIN in the local storage (as is the case when it's the user's first time participating), they will be assigned a random PIN and it will be saved in the browser's local storage. Therefore, if the PIN numbers of two accounts are the same, we know with high probability that they are originating from the same web browser. We note that this may be an underestimate, as if a puppeteer only has a small number of puppets, the chances are lower they would be assigned to the same group. If the puppets are assigned to different groups, the local storage will be set up differently (on a per-group basis). The largest group had 181 participants, so the event of two or more accounts within it being assigned the same random PIN would occur by chance with probability $0.00016$. Therefore, we classify accounts using the same PIN as at least one other account as puppets. We identified at least 38 distinct puppeteers, each having between 2--8 puppets. The number of puppet accounts per puppeteer is likely higher, as their puppets could have been assigned to more than one of the four groups. 

In the next section, we discuss the issue of whether or not the potential puppets might belong to bots. We have limited data to draw upon from Study 2; however, Study 1 has some interesting metrics that we believe serve as indicators that a large number of these accounts had a human behind the keyboard.

\subsection{Could the Puppets Belong to Bots?} 
The distinction between bots and humans in crowdsourcing platforms remains unclear due to the absence of definitive evidence identifying the various types of bots in use. However, in video games, bots take the form of plugins or programs that employ artificial intelligence or automated functions. These bots may assist cheaters by altering user interfaces, modifying game files, or performing tasks that boost their levels, equipment, or in-game currency \cite{lee2016you}. Bots are capable of operating continuously, reacting faster than humanly possible, executing complex tasks in an all-in-one action, or even anticipating future moves \cite{kang2012chatting, chen2008game}. Game managers can often detect such abnormalities in game logs, leading to penalties such as temporary or permanent account bans \cite{kushner2010steamed, thawonmas2008detection}.

In crowdsourcing platforms, while bots might not be as sophisticated as their gaming counterparts, they can still automate tasks. 

Responses from bots tend to be very similar in open-ended questions or exhibit identical patterns in multiple-choice formats. Furthermore, bots typically do not generate unnecessary interaction data, such as additional mouse movements or keystrokes. Time logs of their activities, such as page navigation or the interval between button clicks, are also uniformly consistent.

Study 1 involved more than just a simple questionnaire and password or PIN entry; it involved a number of GUI interactions. As shown in Table \ref{tab:sessions-completion-rates}, most of the participants passed an attention check question ``seven plus three = eight?'' with 5 options from ``Strongly Agree'' to ``Strongly Disagree''. Participants who did not select either ``Strongly Disagree'' or ``Disagree'' were considered as inattentive. However, this alone is not sufficient to distinguish whether the potential puppeteers, for each puppet account, are running automated programs (i.e., bots) or they are a human interacting with the system. 

To better understand whether or not these accounts could be bots, we consider what actions are more ``human'' in our systems, and analyze their occurrence for each potential puppeteer. The results of this analysis are shown in Table \ref{tab:probTable}. We focus our analysis on Study 1 as we have more GUI and user interaction data. 

These events were selected as being a signal of a human behind the screen as they either indicate (a) inefficient behaviours that a bot would not likely adopt, or (b) unique behaviours of each account, which would indicate there is no session reply of a pre-recorded session (these are denoted with a ``*''):

\begin{itemize}
    \item \textbf{Identical Search Term:} True if the accounts use the same search terms (if they searched).
    \item \textbf{No Search:} True if the accounts did not employ the search function.
    \item \textbf{Identical Scrolling*:} True if the accounts share the same scrolling behavior (same number of scroll up and scroll down events).
    \item \textbf{No Scrolling:} True if the accounts have no scrolling activity is present.
    \item \textbf{Default 1st Item Only:} True if only the first default item is selected across the accounts.
    \item \textbf{Identical Patterns*:} True if there are matching answers for all multiple-choice questions.
    \item \textbf{No incorrect Login Attempts:} True if any of the accounts had one or more failed login attempts.
\end{itemize}

While these signals are an indication that there may be an automated program being used by the puppeteer, a single one being true is likely insufficient to classify them as a bot. As shown in Table \ref{tab:probTable}, Pup\_14 is the only puppeteer with more than one of these signals present (No Search and Default 1st Item). It is conceivable that a human would not employ search and always select defaults offered by a system. 

It is also conceivable that puppeteers use automated programs for part of each puppet's tasks (e.g., questionnaires). We note that only one puppeteer shared identical responses between puppets, suggesting human error, boredom, or preferences influenced most puppeteer's answers.

In Study 1, we found very few puppeteers returned for Session 2. For those who returned, they forgot their passwords, leading to multiple failed attempts, often with minor errors like incorrect capitalization or numbers. These varied behaviors, typical of humans, suggest the puppeteers are not automating many of their tasks, despite having identical passwords. Our findings thus indicate a lack of conclusive evidence that these accounts are controlled by bots.

\begin{table*}[t]
\captionsetup{width=.9\textwidth}
\caption{From Study 1, all suspected puppeteers ordered from highest to lowest probability $P(X \geq k)$. The table includes the number of accounts they appear to own (detected from having chosen identical passwords), the probability $p$ of their password in the Pwned Passwords dataset~\cite{PwnedPasswords}, and the resulting probability $P(X \geq k)$ of the password occurring at least $k$ times in our dataset by chance (computed using Equation~\ref{eq:1}). The extremely low values of $P(X \geq k)$ suggest that the identical passwords are unlikely to have occurred by chance.}
\centering
\setlength{\arrayrulewidth}{0.5pt}
\renewcommand{\arraystretch}{1.2}
\resizebox{\textwidth}{!}{%
\begin{tabular}{llll|llll}
\toprule
\textbf{Puppeteers} &
\textbf{\begin{tabular}[c]{@{}l@{}}No. of \\ Accounts ($k$)\end{tabular}} &
\textbf{\begin{tabular}[c]{@{}l@{}}Probability \\ of Success ($p$)\end{tabular}} &
\textbf{\begin{tabular}[c]{@{}l@{}}Probability \\ ($P(X \geq k)$)\end{tabular}} &
\textbf{Puppeteers} &
\textbf{\begin{tabular}[c]{@{}l@{}}No. of \\ Accounts ($k$)\end{tabular}} &
\textbf{\begin{tabular}[c]{@{}l@{}}Probability \\ of Success ($p$)\end{tabular}} &
\textbf{\begin{tabular}[c]{@{}l@{}}Probability \\ ($P(X \geq k)$)\end{tabular}} \\ 
\midrule
Pup\_15 & 3 & $4.12 \times 10^{-5}$ & $1.6 \times 10^{-5}$ & Pup\_9 & 2 & $2.00 \times 10^{-10}$ & $3.3 \times 10^{-16}$ \\
Pup\_20 & 5 & $1.69 \times 10^{-4}$ & $2.4 \times 10^{-6}$ & Pup\_11 & 2 & $2.00 \times 10^{-10}$ & $3.3 \times 10^{-16}$ \\
Pup\_6  & 2 & $5.75 \times 10^{-6}$ & $6.0 \times 10^{-7}$ & Pup\_12 & 2 & $2.00 \times 10^{-10}$ & $3.3 \times 10^{-16}$ \\
Pup\_1  & 2 & $1.50 \times 10^{-6}$ & $1.8 \times 10^{-7}$ & Pup\_13 & 2 & $2.00 \times 10^{-10}$ & $3.3 \times 10^{-16}$ \\
Pup\_4  & 2 & $1.81 \times 10^{-6}$ & $1.6 \times 10^{-7}$ & Pup\_28 & 8 & $4.54 \times 10^{-6}$ & $1.2 \times 10^{-19}$ \\
Pup\_7  & 2 & $1.30 \times 10^{-6}$ & $1.1 \times 10^{-7}$ & Pup\_14 & 3 & $2.00 \times 10^{-10}$ & $1.8 \times 10^{-23}$ \\
Pup\_5  & 2 & $1.15 \times 10^{-6}$ & $7.6 \times 10^{-8}$ & Pup\_16 & 3 & $2.00 \times 10^{-10}$ & $1.8 \times 10^{-23}$ \\
Pup\_18 & 4 & $1.35 \times 10^{-5}$ & $1.1 \times 10^{-8}$ & Pup\_30 & 19 & $1.24 \times 10^{-3}$ & $2.3 \times 10^{-26}$ \\
Pup\_10 & 2 & $1.20 \times 10^{-8}$ & $1.2 \times 10^{-12}$ & Pup\_22 & 5 & $3.82 \times 10^{-8}$ & $1.0 \times 10^{-27}$ \\
Pup\_24 & 7 & $7.40 \times 10^{-4}$ & $2.2 \times 10^{-12}$ & Pup\_25 & 7 & $1.31 \times 10^{-6}$ & $3.2 \times 10^{-29}$ \\
Pup\_17 & 4 & $6.84 \times 10^{-7}$ & $5.0 \times 10^{-13}$ & Pup\_19 & 4 & $2.00 \times 10^{-10}$ & $1.2 \times 10^{-31}$ \\
Pup\_27 & 8 & $1.99 \times 10^{-5}$ & $1.1 \times 10^{-13}$ & Pup\_23 & 5 & $8.10 \times 10^{-9}$ & $1.2 \times 10^{-32}$ \\
Pup\_3  & 2 & $3.20 \times 10^{-9}$ & $8.4 \times 10^{-14}$ & Pup\_26 & 7 & $1.40 \times 10^{-9}$ & $1.7 \times 10^{-50}$ \\
Pup\_21 & 5 & $3.22 \times 10^{-6}$ & $1.5 \times 10^{-17}$ & Pup\_29 & 13 & $2.00 \times 10^{-10}$ & $2.1 \times 10^{-74}$ \\
Pup\_2  & 2 & $2.00 \times 10^{-10}$ & $3.3 \times 10^{-16}$ & Pup\_31 & 57 & $3.90 \times 10^{-9}$ & $4.5 \times 10^{-279}$ \\
Pup\_8  & 2 & $2.00 \times 10^{-10}$ & $3.3 \times 10^{-16}$ &        &   &   &   \\
\bottomrule
\end{tabular}%
}
\caption{Summary of puppeteer detection probabilities.}
\label{tab:probTable}
\end{table*}

\subsection{How Many Accounts are Controlled by Puppeteers?}

Here we present the resulting number of suspicious accounts for each study. For Study 1, we broke them down into inattentive accounts (who failed the aforementioned attention check question) and puppeteers. 
Among 558 participants, we observed 193 potential puppet accounts (35.1\%). See Table \ref{tab:sessions-completion-rates} for full details. For Study 2, among 698 participants, we identified 394 (56.5\%) potential puppet accounts (see Table \ref{tab:sessions-completion-rates-2} for full details). As this study lacked a Session 1 attention check, it does not have this row, but we find it noteworthy that the combined rate of inattentive + puppet accounts for Study 1 is 57\%, comparable to the number of puppet accounts identified in Study 2 (56.5\%).  

\section{Strategies for Detecting Fraudulent Activities}
We discuss a number of strategies to detect both bots and puppeteers. For completeness, we describe some existing methods that continue to offer some benefits (e.g., much of the bot detection strategies described). To the best of our knowledge, the puppeteer detection strategies we discuss are novel. We hope to have fruitful discussion about these and possibly other detection strategies at NSPW.

\subsection{Bot Detection Strategies}

Bot detection serves as the foundational layer in protecting data integrity by identifying automated programs. However, the distinction between bots and humans in crowdsourcing platforms remains unclear due to the absence of definitive evidence identifying the various types of bots in use. However, in the realm of gaming, bots represent a significant issue, causing distress for developers and, in extreme cases, leading to the bankruptcy of companies \cite{kang2012chatting}. 
Bots are capable of operating continuously, reacting faster than humanly possible, executing complex tasks in an all-in-one action, or even anticipating future moves \cite{kang2012chatting, chen2008game}. Game managers can often detect such abnormalities in game logs, leading to penalties such as temporary or permanent account bans \cite{kushner2010steamed, thawonmas2008detection}. This section discusses several methods in detecting bots and outlines how puppeteers might evade these measures. 

In crowdsourcing platforms, while bots might not be as sophisticated as their gaming counterparts, they can still automate tasks. For instance, they may memorize and replicate the coordinates of buttons and input fields required in questionnaires. With the advancement of AI, smarter bots might soon be capable of scraping questionnaire content, feeding it into generative AI models, and generating plausible answers automatically.
\subsubsection{Bot Types} 
We have categorized three distinct types of bots based on their characteristics and capabilities:
\begin{itemize}
    \item \textbf{Replay Bots (RB)} track screen coordinates only and replay events. They perform repetitive actions based on the recorded coordinates, making them suitable for straightforward tasks that do not require context or complex decision-making. 
\end{itemize}
\begin{itemize}
    \item \textbf{Smart Bots (SB)} are even more advanced. They scan each clicked item, trigger the necessary events, such as filling forms with preset random text, selecting random options, clicking ``Next'' buttons, and complete the survey by triggering the ``Submit'' functions. These bots can handle more complex interactions by understanding and responding to the context of each task. 
\end{itemize}
\begin{itemize}
    \item \textbf{Gen-AI Bots (GB)} represent the cutting edge of automation. By integrating web scraping tools such as Selenium \cite{Selenium} or Puppeteer \cite{PuppeteerAPI} with generative AI tools like ChatGPT \cite{openAIAPI}, it is possible to create a survey bot capable of performing tasks with minimal human intervention. For instance, Selenium can scrape a survey's front-end html code, extract options and buttons, and trigger events. When handling free-form questions, instead of generating random or repetitive content, the bot requests dynamic answers using generative AI.
\end{itemize}

\subsubsection{Attention Questions.} Attention Questions have been employed by researchers for decades \cite{berinsky2014separating,meade2012identifying,woods2006careless}. Requiring correct responses to specific questions ensures participants are engaged and not automated scripts. Bots might be programmed to answer these, but incorrect or inconsistent answers can also be flagged as puppeteers. No matter they are bots or puppeteers, failure to do so will disqualify them from participating, and their data will be excluded from the studies. Although this is a general method to detect most bots (except Gen-AI), most of the puppeteers can pass attention questions.%without clues will be revealed \cite{Hauser2022CloudResearch, kennedy_clifford_burleigh_waggoner_jewell_winter_2020}.

\subsubsection{Free-form Questions.} Identical answers to free-form questions are rare to see. However, simple, fixed free-form questions may lead to the same answers, such as ``good'' or ``ok''.  If one-word, irrelevant, or random responses appear at the same time across different questions, it can be used as indicators of bots, or low-quality responses from puppeteers or inattentive workers \cite{chmielewski2020mturk,kennedy2020assessing}. This method can detect Replay Bots.

To further enhance this method, we suggest that researchers integrate data collection expected to be unique and follow a known probability distribution, such as passwords, longer personal responses, or other forms of input that are inherently varied. This approach can make it more challenging for automated bots to generate authentic-looking responses, as it presents another hurdle them to mimic the natural variation expected in human responses that adhere to specific probability distributions.

\subsubsection{Time Differences} 
Time difference in questionnaire responses is a metric for detecting bots as bots can respond much faster and more uniformly than humans. Unlike humans, who exhibit variability in response times due to factors like question complexity and reading comprehension, bots often show unnaturally quick and consistent response times across all questions. Initial response times that are significantly shorter than typical human response times, uniformly fast answers, and predictable intervals between responses can all indicate automated behavior. By comparing these time metrics against established human benchmarks, abnormal responses that suggest non-human interactions can be identified. This method can detect all types of bot. However, bots can add dynamic/random time delays to mimic legitimate human behaviors to evade this detection. 

\subsubsection{Question Patterns.} This test assesses the logical consistency and uniqueness of the user's responses to multiple choice, demographic, or likert scale questions, distinguishing between genuine user input and patterned, possibly automated responses. For example, some participants had provided the same responses with specific patterns, such as ``111111'' or ``121212'', this can be used as an indicator to determine either they are automated responses from replay bots or puppeteers who want to rush studies.

\subsubsection{Machine Information}

A participant's device always contains unique information, Such as IP address, geolocations, and browser plug-ins. However, due to certain privacy regulations, sensitive data such as IP address cannot be easily stored and processed at researcher's server. Thus, we suggest researchers use non-sensitive unique identifiers to determine whether a participant's machine information is either unique or identical to others.

One method to achieve this is through generating unique browser fingerprints. This technique involves the use of software libraries to create a unique identifier named ``fingerprint'' based on a combination of browser information and settings. For instance, a commonly used privacy-aware library \cite{FingerPrintJS} provides a way to generate such fingerprints anonymously, ensuring that they cannot be traced back to individuals. These fingerprints are useful for identifying bots and puppeteers, even if the browser is in incognito/private mode. However, the uniqueness of each fingerprint and the likelihood of any two being identical should be assessed carefully, often in combination with additional identification methods.

Another method involves analyzing data stored locally on a user's browser, such as cookies or local storage. These tools can store vital information, including session data and unique values which are typically unlikely to be identical accidentally among different users. The use of local storage might include data that remains persistent until explicitly cleared, whereas cookies could be automatically reset or deleted, especially in private/incognito mode. This could affect the reliability of identifying puppeteers or bots under certain conditions.

Researchers should calculate the likelihood of identifier collisions to estimate the probability of such occurrences, to reduce the chance of false positive matches. This method can be used to detect all types of bots and puppeteers; however it is possible for it to be evaded if the bot-master or puppeteer takes extra precautions.

\subsubsection{Increasing Cost Using Text Images}\label{tti}
Most of the bot detection methods discussed above can detect Replay Bots and Smart Bots; however Gen-AI Bots remain a challenge. One potential method to increase the difficulty and cost of data extraction is to transfer textual content into images, leveraging the increased computational resources required to process images compared to text, to raise their costs. This method involves replacing textual elements with images that visually represent the text, requiring bots to employ Optical Character Recognition (OCR), which is computationally expensive and time-consuming compared to parsing HTML text, and can potentially introduce errors. This added complexity means bots must handle image downloads, image-to-text conversion, and error correction, increasing the processing difficulty and resource usage. As a result, a full survey could contain thousands of images, making the operational costs for bots much higher than the potential benefits, thereby discouraging fraudulent activities.  However, implementing this strategy requires balancing usability for human users with deterrence for bots, ensuring high-quality, readable images, optimizing image sizes for performance. 

\subsection{Puppeteer Detection Strategies}

Comparing to bot detection, identifying a puppeteer requires more advanced detection strategies when they are able to evade most of the above bot detection methods. This seems especially likely as their awareness of detection methods increases, and they adapt their strategies accordingly. 

\subsubsection{Behavioral Anomaly Detection and Clustering} 

Behavioral patterns such as mouse movements and keyboard dynamics can be used to detect anomalies that could indicate bots (e.g., extremely fast movements). Patterns in these behaviours may also be useful to match one participant to others in the same crowdsourced dataset, indicating a common puppeteer behind each. Behavioural data can be collected for each participant based on typing speed, keystroke patterns, and mouse movement trajectories. Such techniques have been attempted for authentication, to match a user's patterns to a trained template created through registration \cite{revett2008survey, shen2012user, ma2016kind,mondal2017study}. These methods may be useful for our puppet detection purpose by using them to facilitate clustering respondents into potential puppeteers. This approach would use this data for grouping participants together, rather than matching to a pre-trained template, so it would not authenticate or identify a participant.

\subsubsection{Implicit Learning Tests} \label{ilt}

Integrating an implicit learning task such as contextual cueing can enrich puppet detection strategies. The goal of this technique is to determine whether a participant has ever participated in the study before. 
The task requires participants to identify a unique character within a 2D display of characters, a process typically repeated 4--6 times to ensure learning \cite{chun1998contextual}. We assume that each unique crowdsourcing study will use its own unique 2D display. Upon receiving training for the first time, genuine participants would be expected to show progressively faster search times with each repetition, demonstrating a natural learning curve. In contrast, if the participant has received the training already through using a different account, they are likely to maintain fast search times from the beginning, without the gradual improvement indicative of learning for the first time. 
Additionally, this method can prove effective in identifying bots as they are likely to have faster times and/or lack the human learning curve. As an added benefit, this method can also deter Gen-AI bots, as this task (at least currently) cannot be solved with ChatGPT 4 \cite{openAIAPI}.

\subsubsection{Dynamic Questions}
The static nature of question answers renders them vulnerable to manipulation, especially after several rounds of tasks when fraudulent workers or bots have identified the correct responses and are ready to share them. Under this situation, we suggest researchers implement dynamic positions and randomized contexts to tackle this issue.

\begin{itemize}
    \item \textbf{{Dynamic Positions:}} To ensure the uniqueness of question presentation and avoid recognizable patterns, we employ a strategy where the positions of multiple choice questions are randomized for each participant. This approach prevents any two participants from viewing questions in the same sequence, thereby minimizing the chance of general patterns emerging. Additionally, by analyzing responses for any suspicious or identical answering patterns, we can detect participants who may not be considering the context of the questions. This randomization is efficiently achieved by generating unique seeds for each participant using JavaScript.
\end{itemize}

\begin{figure}[tb]
\centering
  \includegraphics[width=0.47\textwidth]{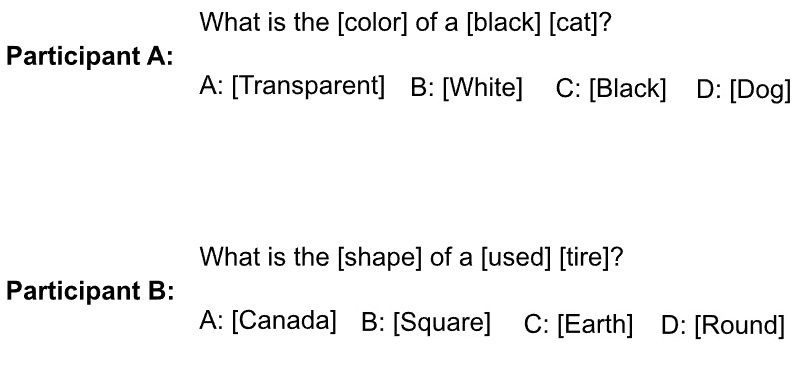} 
  \caption{Simple example of a dynamic multiple choice question. Text in brackets are dynamic words inserted on the fly from an online or local database. The dynamic text will not show different font styles to make them unattractive. 
  }
  \label{img:cat}
\end{figure}

\begin{itemize}
    \item \textbf{Randomized Context:} Introducing variability in question context can frustrate puppeteers who rely on repetitive answers. Each participant sees personalized free-form questions when the context of a free-form question can be rewritten using NLP techniques to have totally different answers. For example, participant A sees “What is the color of a banana?” Participant B sees “What is the color of a cherry?” They should provide different answers for their questions. Unlike reCAPTCHA, which relies solely on user mouse interaction (such as clicks or drags), this approach introduces more variations in question content, making manipulation more challenging. Additionally, reCAPTCHA's purpose is often more apparent compared to this method. Figure \ref{img:cat} shows some examples.
\end{itemize}

\section{Discussion}

\subsection{Impact of Puppeteers}

Puppeteers represent a significant, yet under explored, challenge in crowdsourced data collection. The issue arises when single participants manually manage multiple accounts, easily completing straightforward tasks and attention checks. Researchers lacking advanced web programming expertise may either ignore or underestimate the serious impact of such deceptive activities. Without effective detection to allow removal of noisy puppet-generated data, it can lead to:

\begin{table}[ht]
\centering
\caption{Comparison of various results from a study of an approach to nudge users to create stronger passwords \cite{wang2023pixi}, with and without puppet noise.  The comparison indicates that the password security improvement changes depending on whether puppet noise is filtered from the dataset. Note the p-values were considered after Holm-Bonferroni multiple-test correction ($\alpha^{\prime}_{(1)}=0.0167$, $\alpha^{\prime}_{(2)}=0.025$, $\alpha^{\prime}_{(3)}=0.05$). Password strength was measured using the CMU Password Guessability Service \cite{cmu} as described in \cite{wang2023pixi}.}
\begin{tabular}{cccc}
\hline
\textbf{\begin{tabular}[c]{@{}c@{}}Security \\ Metric\end{tabular}}                                                            & \textbf{Results}            & \textbf{\begin{tabular}[c]{@{}c@{}}Clean \\ (No Puppets)\end{tabular}}                           & \textbf{\begin{tabular}[c]{@{}c@{}}Noisy \\ (Incl. Puppets)\end{tabular}}                          \\ \hline
\multirow{3}{*}{\begin{tabular}[c]{@{}c@{}}Password \\ length\end{tabular}}   & Rejected ${\mathcal{H}}_{0}$ & Yes                         & No                             \\
                                                                              & P value                     & \textbf{0.002*} $< \alpha^{\prime}_{(1)}$          & 0.024 $< \alpha^{\prime}_{(1)}$           \\
                                                                              & Effect size                 & Large (0.44)                & Small (0.0202)                 \\ \hline
\multirow{3}{*}{\begin{tabular}[c]{@{}c@{}}Password \\ Strength\end{tabular}} & Rejected ${\mathcal{H}}_{0}$ & Yes                         & No                             \\
                                                                              & P value                     & \textbf{0.002*} $< \alpha^{\prime}_{(2)}$          & 0.466 $> \alpha^{\prime}_{(2)}$           \\
                                                                              & Effect size                 & Medium (0.187)              & -                              \\ \hline
\multirow{3}{*}{\begin{tabular}[c]{@{}c@{}}Zxcvbn \\ Score\end{tabular}}      & Rejected ${\mathcal{H}}_{0}$ & Yes                         & No                             \\
                                                                              & P value                     & \textbf{0.022*} $< \alpha^{\prime}_{(3)}$          & 0.321 $> \alpha^{\prime}_{(3)}$           \\
                                                                              & Effect size                 & Medium (0.32)               & -                              \\ \hline
\end{tabular}
\label{tab:new-table}
\end{table}

\begin{itemize}
    \item \textbf{Null results (false negatives)} for well designed studies with appropriate control groups. Such false negatives can prevent research efforts from proceeding down useful paths. For example, in our nudging stronger password creation study \cite{wang2023pixi}, the security results would have been null if puppets were not filtered out (see Table \ref{tab:new-table}). Our analysis serves as evidence to the noise that such puppets contribute to the data. 

\item \textbf{False positives} occur when studies incorrectly indicate the presence of an effect that does not actually exist. For example, in well-designed studies with appropriate control groups, this might happen if the noise created by puppets obscures differences between groups, leading them to appear similar or equivalent. This may result in incorrect conclusions that the groups are not impacted by the experimental conditions.

\item \textbf{Inaccurate or misleading data} can arise in exploratory studies that lack control groups. Absolute values collected from crowdsourced populations may differ from those collected from the broader population, not only due to differences in the population but also due to fraudulent activities.
    
\item \textbf{Reduced diversity of the participant pool. }

\item \textbf{Increased costs.} If proactive methods to detect (in order to reject) puppets early in the job's workflow are not performed, the costs of using crowdsourcing becomes much higher. 
\end{itemize}

\subsection{Why do Puppeteers Exist?}

At a high level, puppeteers and bot-masters may seem similar, but they act differently. Puppeteers  may employ more traditional methods due to factors such as time constraints, labor costs, technological limitations, or personal preferences. On the other hand, bot-masters are mechanized workers whose initial investment in acquiring or developing fully or partially automated programs is higher compared to puppeteers. While new technologies may encourage puppeteers to adopt state-of-the-art programs for efficiency, this transition is not always straightforward.
The output efficiency of bot-masters is remarkable, but it comes at the expense of high initial implementation costs and potential challenges in adapting to future tasks. 

To illustrate this, consider the analogy of agriculture: each crowdsourcing task is similar to an independent farmland, with various conditions like plains, ridges, or rivers.
Puppeteers then resemble traditional farm workers who may hire themselves or individuals from regions with lower labor costs to complete jobs. They have flexibility to adapt to different conditions but have to work at a slower speed.
Depending on the size of the job, the return on investment may not be worthwhile to automate all tasks. As a consequence, similar to longstanding practices in other areas such as agriculture, there will likely always be a presence of traditional workers whose roles remain irreplaceable in the foreseeable future.

\subsection{Security Through Obscurity}
Some researchers employ security through obscurity by keeping their fraud detection methods and system details confidential. This strategy aims to prevent fraudsters adapting to detection techniques by withholding information about how these methods operate.

However, relying solely on obscurity has significant limitations. Fraudsters may eventually uncover hidden mechanisms through experimentation or information sharing. Moreover, keeping fraud detection methods secret can conflict with open science principles, which advocate for transparency and reproducibility in research. By not disclosing these methods, it becomes challenging for other researchers to reproduce studies, verify results, or build upon the work.

\subsection{Forecasting the Arms Race}

Puppeteers will likely develop new strategies to evade detection. It seems likely that the bot detection methods outlined will be evaded by activities like VPS/VPN (already in use), local storage and cookie flushing between sessions, and employing AI techniques to automate questionnaire responses. Systems will need to stay one step ahead of current AI capabilities to combat this for simple questionnaire-based studies. At this point in time, converting text to a large number of small images for OCR to process (recall Section \ref{tti}) may increase the cost such that it is no longer worth it for AI bots to exploit crowdsourced tasks. Keeping the cost of crowdsourcing fraud in mind, and deploying mitigation strategies such that they increase such cost, may be the most effective strategy. Crowdsourcing for novel user interface studies may have an advantage as the economic incentives may not be present for a puppeteer to create a custom AI solution to complete a study involving a new piece of software and its unique activities. Unique approaches such as the Implicit Learning Task of Section \ref{ilt} can also require a custom AI solution. However, it is important to keep in mind that a puppeteer may automate only some of the parts of a crowdsourced task in order to complete the study with the most accounts in the least time. Therefore, it may be important to deploy detection strategies for every part of the crowdsourced task (not only the pre-screening). 

\subsection{Ethical Considerations for Puppet Detection}
Various compliance requirements, such as Research Ethics Board (REB) in Canada \cite{REB}, the Institutional Review Board (IRB) \cite{grady2015institutional} in the United States, or the European Network of Research Ethics Committees (EUREC) \cite{EUREC} introduce another layer of complexity. Researchers are normally expected to ensure participants receive incentives regardless of the quality of data they provide, and that their payment meets or exceeds the local minimum wage. The best way to handle this is typically by employing detection methods in a pre-screening questionnaire. Once a participant passes a quick pre-screen, indicating they are not a bot or puppet, they can proceed with the study to receive compensation. Detection methods that require more data than can be reliably collected in the pre-screen could be employed and used to stop the study part way through in order to conserve partial compensation resources.

\subsection{Puppeteer Detection Costs}

Implementing and maintaining effective detection and prevention strategies against puppeteers may bring significant costs for researchers. These costs include time, financial resources, and the need for specialized expertise. Researchers must invest in developing or acquiring detection systems, continuously update these systems to keep up with evolving tactics, and allocate time for manual verification processes. Smaller institutions or individual researchers may find these requirements particularly burdensome, potentially limiting their ability to conduct robust crowdsourcing studies.

\subsection{Platform Level - Increasing Cost for Fraudsters}
An effective method to combat bots and puppeteers is to raise their operational costs, making fraud too expensive to pursue.

Crowdsourcing platforms could introduce a deposit system, requiring users to submit a deposit before participating in tasks. If a user is consistently flagged for providing low-quality data—whether detected by the platform or reported by researchers—a portion of their deposit would be deducted. This approach creates a direct financial deterrent for fraudsters, as they risk losing money for submitting poor-quality data. However, careful consideration must be given to designing comprehensive policies that protect legitimate users from unjust penalties, ensuring fairness while discouraging fraudulent behavior.

\subsection{Raising Awareness of the Puppeteers Issue}
It will be important to raise awareness within the research community regarding puppeteers. Efforts will likely start with papers such as this being discussed in conferences and workshops. As awareness spreads, hopefully reviewers will advocate for the inclusion of puppeteer detection as a method to ensure crowdsourced data quality. To facilitate this, the community should develop free resources, such as guidelines, tutorials, and toolkits, which clearly outline proposed strategies for detecting puppeteer behaviors. These resources could aim to help researchers with the necessary tools to effectively identify and mitigate these issues in their data collection. To gain access to such resources, it may be wise to have an application process to ensure the resources are not exploited by the puppeteers themselves to help them evade detection. 

\subsection{Are Puppets a Platform Problem?}
Crowdsourcing platforms play a crucial role in helping researchers collect clean and accurate data. However, the presence of bots and puppeteers presents a significant challenge to the integrity of the data collected. While researchers must carefully sanitize their datasets to remove noisy data, the responsibility for detecting and managing fraudulent activities, such as bots or puppets, should not fall solely on them.

Platforms, which have full control over the infrastructure and user management systems, are in a strong position to implement effective bot detection and account verification measures. However, they may lack financial motivation to actively eliminate these fraudulent activities, as a higher volume of workers translates to more revenue from task completion fees. This creates a conflict of interest where platforms benefit from increased participation, even if some of that participation is fraudulent.

Nevertheless, there is a clear and compelling research interest in ensuring data integrity. Poor data quality resulting from the activities of puppeteers and bots not only undermines individual research projects but also threatens the reputation of the platforms as reliable research tools.

As a result, ensuring the integrity of data collected on crowdsourcing platforms will likely be a shared responsibility. While researchers should implement detection strategies, platforms must also be held accountable for providing a secure and reliable environment that prioritizes data quality over participation volume.

\section{Future Work}
We found this puppeteers issue on Amazon MTurk via two user studies that were designed for other research purposes. Future studies should be designed to specifically explore puppeteers in more detail, and investigate the extent to which this issue exists on other crowdsourcing platforms (e.g., Prolific). While newer platforms are often thought to be better, it is crucial to recognize that they might also suffer from similar issues. Prolific is increasingly favored by researchers due to perceived higher quality, so it is essential to determine if similar issues are present. Investigating whether the puppeteer problem persists on Prolific or other newer platforms could provide valuable insights into the reliability of these platforms. Additionally, the detection methods we suggest in this paper should be studied to determine their efficacy. Since many of these methods are still in the prototype stage, it is crucial to assess their effectiveness in real-world applications. It is possible that some subset of detection methods is most useful and/or cost-effective. Moreover, raising the operational costs for fraudsters through techniques such as text-to-image conversion could be a useful approach to investigate in future research.

\section{Conclusions}
Our work unveils a significant challenge: the prevalence of puppeteers accounts, leading to fraudulent data entries originating from the same entity. This revelation, drawn from secondary data analysis of two studies, signals a pressing need for researchers to move beyond conventional validation methods like attention checks, which these fraudulent workers easily bypass. To combat this concern, we advocate for the implementation of a layered approach to data validation, combining traditional methods with novel techniques that are designed to target puppeteers. By deploying these strategies, researchers can protect the integrity of their data against such fraudulent participants and bots, ensuring the continued utility of Amazon MTurk and similar crowdsourcing platforms for higher-quality research outcomes.

Looking ahead, it is crucial that future research remains adaptable. There is no denying that puppeteers will keep evaluating their techniques, leveraging advancements in AI and automation to evade detection. Researchers must develop and update countermeasures that are equally dynamic and capable of evolving alongside these threats. 

\section*{Acknowledgment}
The authors would like to thank the NSPW peer reviewers for their
feedback. This research was supported by the Natural Sciences and Engineering Research Council of Canada (NSERC).

\bibliography{sample-base}

\end{document}

%% file: Study1.tex
\begin{table*}[htb]
\captionsetup{width=.9\textwidth}
\caption{Number of MTurk users for Study 1 who participated, and were found to be puppets, inattentive, or valid across conditions.}
\centering
\begin{tabular}{l>{\centering}p{2cm}>{\centering}p{2cm}>{\centering}p{5.2cm}>{\centering\arraybackslash}c}
\toprule
\textbf{} & \textbf{Group \#1} & \textbf{Group \#2} & \textbf{Group \#3} & \textbf{Overall} \\
\cmidrule(lr){2-5}
Participants & 181 & 185 & 192 & 558 (100\%) \\
\midrule
Puppets & 76 & 53 & 64 & 193 (34.6\%) \\ 
Inattentive & 34 & 49 & 44 & 127 (22.7\%) \\ 
\midrule
Valid Workers & 71 & 83 & 84 & 238 (42.7\%) \\ 
\bottomrule
\end{tabular}
\label{tab:sessions-completion-rates}
\end{table*}

%% file: Study2.tex
\begin{table*}[htb]
\captionsetup{width=.9\textwidth}
\caption{Number of MTurk users for Study 2 who participated, categorized as puppets or valid workers across conditions.}
\centering
\begin{tabular}{l>{\centering}p{2cm}>{\centering}p{2cm}>{\centering}p{2cm}>{\centering}p{3.2cm}>{\centering\arraybackslash}c}
\toprule
\textbf{} & \textbf{Group \#1} & \textbf{Group \#2} & \textbf{Group \#3} & \textbf{Group \#4} & \textbf{Overall} \\
\cmidrule(lr){2-6}
Participants & 167 & 181 & 178 & 172 & 698 (100\%) \\
\midrule
Puppets & 88 & 97 & 91 & 108 & 384 (55\%) \\ 
\midrule
Valid Workers & 77 & 84 & 87 & 69 & 317 (45\%) \\
\bottomrule
\end{tabular}
\label{tab:sessions-completion-rates-2}
\end{table*}